\documentclass[graybox]{svmult}

\usepackage[breaklinks = true,colorlinks = true,linkcolor = blue,citecolor = blue]{hyperref}

\usepackage{type1cm}        
\usepackage{makeidx}         
\usepackage{graphicx}        
\usepackage{multicol}        
\usepackage[bottom]{footmisc}

\usepackage{newtxtext}       %
\usepackage{newtxmath}       

\newcommand{\calL}{\mathcal{L}}

\newcommand{\bnabla}{\boldsymbol{\nabla}}
\newcommand{\bgamma}{\boldsymbol{\gamma}}
\newcommand{\ba}{\boldsymbol{a}}
\newcommand{\bbm}{\boldsymbol{m}}
\newcommand{\bp}{\boldsymbol{p}}
\newcommand{\bx}{\boldsymbol{x}}
\newcommand{\bA}{\boldsymbol{A}}
\newcommand{\bB}{\boldsymbol{B}}
\newcommand{\bD}{\boldsymbol{D}}
\newcommand{\bE}{\boldsymbol{E}}

\newcommand{\bJ}{\boldsymbol{J}}
\newcommand{\bL}{\boldsymbol{L}}
\newcommand{\bS}{\boldsymbol{S}}

\makeindex             

\begin{document}

\title*{Relativistic decomposition of the orbital and the spin angular
momentum in chiral physics and Feynman's angular momentum paradox}
\titlerunning{OAM/SAM decomposition} 
\author{Kenji Fukushima and Shi Pu}
\institute{Kenji Fukushima \at Department of Physics, The University
  of Tokyo, 7-3-1 Hongo, Bunkyo-ku, Tokyo 113-0033, Japan, \email{fuku@nt.phys.s.u-tokyo.ac.jp}
\and Shi Pu \at Department of Modern Physics, University of Science
and Technology of China, Hefei 230026, China, \email{shipu@ustc.edu.cn}}
%
%
\maketitle

\abstract*{
  Over recent years we have witnessed tremendous progresses in our 
  understanding on the angular momentum decomposition.  In the context 
  of the proton spin problem in high energy processes the angular 
  momentum decomposition by Jaffe and Manohar, which is based on the 
  canonical definition, and the alternative by Ji, which is based on 
  the Belinfante improved one, have been revisited under light shed by 
  Chen \textit{et al.} leading to seminal works by Hatta, Wakamatsu, 
  Leader, etc.  In chiral physics as exemplified by the chiral
  vortical effect and applications to the relativistic nucleus-nucleus
  collisions, sometimes referred to as a relativistic extension of the
  Barnett and the Einstein--de~Haas effects, such arguments of the
  angular momentum decomposition would be of crucial importance.  We
  pay our special attention to the fermionic part in the canonical
  and the Belinfante conventions and discuss a difference between
  them, which is reminiscent of a classical example of Feynman's
  angular momentum paradox.  We point out its possible relevance to
  early-time dynamics in the nucleus-nucleus collisions, resulting in
  excess by the electromagnetic angular momentum.
  }

\abstract{
  Over recent years we have witnessed tremendous progresses in our 
  understanding on the angular momentum decomposition.  In the context 
  of the proton spin problem in high energy processes the angular 
  momentum decomposition by Jaffe and Manohar, which is based on the 
  canonical definition, and the alternative by Ji, which is based on 
  the Belinfante improved one, have been revisited under light shed by 
  Chen \textit{et al.} leading to seminal works by Hatta, Wakamatsu, 
  Leader, etc.  In chiral physics as exemplified by the chiral
  vortical effect and applications to the relativistic nucleus-nucleus
  collisions, sometimes referred to as a relativistic extension of the
  Barnett and the Einstein--de~Haas effects, such arguments of the
  angular momentum decomposition would be of crucial importance.  We
  pay our special attention to the fermionic part in the canonical
  and the Belinfante conventions and discuss a difference between
  them, which is reminiscent of a classical example of Feynman's
  angular momentum paradox.  We point out its possible relevance to
  early-time dynamics in the nucleus-nucleus collisions, resulting in
  excess by the electromagnetic angular momentum.
}

\section{Prologue}
\label{sec:prologue}

Some time ago we, Fukushima and Pu, together with our bright
colleague, Zebin Qiu, published a paper~\cite{Fukushima:2018osn} on a
relativistic extension of the Barnett effect~\cite{RevModPhys.7.129}
in the context of chiral materials.  Our results are beautiful and
robust, we believe, but at the same time, we had to overcome many
conceptual confusions.  We are 100\% sure about our calculations,
results, and conclusions, but we were unable to find 100\% unshakable
justification to our spin identification.  We could not remove theoretical
uncertainty to extract the orbital angular momentum (OAM) and the spin
angular momentum (SAM) out of the total angular momentum that is
conserved.  We adopted the most natural assumption, meanwhile we
studied many preceding works;  for example, we found
Ref.~\cite{PhysRevLett.118.114802} that makes a surprising assertion
of the existence of individually conserved OAM and SAM derived from
the Dirac equation.  The more we studied, the more confusions we were
falling into.  The present contribution is not an answer to
controversies, but more like a note of what we have understood so
far, and some our own thoughts based on them.  Actually in
Ref.~\cite{Fukushima:2018osn} we posed an important question of
how to represent the Barnett effect in chiral hydrodynamics, but in
the present article we will not mention this.  We will report our
progresses on hydrodynamics with OAM and SAM somewhere else hopefully
soon, and the present article is focused on the field theoretical
descriptions.

\section{Basics -- Angular momenta in an Abelian gauge theory}
\label{sec:conserved}

In non-relativistic and classical theories the spin is not a dynamical
variable;  spin-up and spin-down electrons are treated as distinct
species and the total spin is conserved unless interactions allow for
spin unbalanced processes.  Dirac successfully generalized an equation
proposed by Pauli, who first postulated such internal doubling, into a
fully relativistic formulation.  Eventually Majorana and other
physicists realized the usage of Cartan's spinors.  Today, even
undergraduate students are familiar with tensors and spinors according
to the representation theory of Lorentz symmetry.  In contemporary
physics symmetries and associated conserved quantities play essential
roles. This article mainly addresses the angular momentum and the
spin.  Readers interested in the history of the spin are invited to
consult a very nice book, \textit{The Story of Spin}, by
Sin-itiro~Tomonaga (see Ref.~\cite{tomonaga1998story} for an English
translated version).

To begin with, we shall summarize some textbook knowledge about
various assignments of angular momenta.  Lorentz symmetry is
characterized by the following transformation,
\begin{equation}
  x^\mu \;\to\; x^{\prime \mu} = \Lambda^\mu_{\;\nu} x^\nu
  = (\delta^\mu_{\;\nu} + \epsilon^\mu_{\;\nu}) x^\nu\,,
  \label{eq:Lorentz}
\end{equation}
where $\Lambda_{\mu\nu}$ and infinitesimal $\epsilon_{\mu\nu}$ are
antisymmetric tensors.  Let us take a simple Abelian gauge theory
defined by the following Lagrangian density,
\begin{equation}
  \calL = \bar{\psi}(i\gamma^\mu D_\mu - m)\psi
  -\frac{1}{4} F^{\mu\nu}F_{\mu\nu}
\end{equation}
with the covariant derivative, $D_\mu\equiv \partial_\mu +ieA_\mu$, and
the field strength tensor,
$F^{\mu\nu}\equiv \partial^\mu A^\nu-\partial^\nu A^\mu$.  This theory
involves vector and spinor fields which transform together with
Eq.~\eqref{eq:Lorentz} as
\begin{align}
  A^\mu(x) & ~\to~
    A^{\prime \mu}(x) = \Lambda^\mu_{\;\nu} A^\nu(\Lambda^{-1} x) \,,\\
  \psi(x) & ~\to~
    \psi'(x) = \Lambda_{\frac{1}{2}} \psi(\Lambda^{-1} x) \,,
\end{align}
where
$\Lambda_{\frac{1}{2}} = \boldsymbol{1} - \frac{i}{2} \epsilon_{\mu\nu} \Sigma^{\mu\nu}$
with
$\Sigma^{\mu\nu} \equiv \frac{i}{4}[\gamma^\mu,\gamma^\nu]$.
Thus, for an infinitesimal transformation, the fields change as
$A^\alpha(x)\to A^\alpha(x)+\frac{1}{2}\epsilon_{\mu\nu}\Delta A^{\mu\nu\alpha}(x)$ and
$\psi(x)\to\psi(x)+\frac{1}{2}\epsilon_{\mu\nu}\Delta\psi^{\mu\nu}(x)$
(where we put $\frac{1}{2}$ for antisymmetrization)
with
\begin{align}
  \Delta A^{\mu\nu\alpha}(x) &=
    \Bigl[ (x^\mu\partial^\nu - x^\nu\partial^\mu) g^{\alpha\beta}
    +(g^{\mu\alpha} g^{\nu\beta}-g^{\nu\alpha}g^{\mu\beta}) \Bigr] A_\beta(x)\,,\\
  \Delta\psi^{\mu\nu}(x) &= \bigl( x^\mu\partial^\nu - x^\nu\partial^\mu
  - i\Sigma^{\mu\nu} \bigr) \psi(x)\,.
\end{align}
Now we can compute the N\"{o}ther current.  From the gauge part we
find,
\begin{equation}
  J_A^{\lambda\mu\nu} = \frac{\partial\calL}{\partial(\partial_\lambda A^\alpha)}
  \Delta A^{\mu\nu\alpha} =
  -F^\lambda_{\;\alpha} (x^\mu\partial^\nu - x^\nu\partial^\mu) A^\alpha
  -F^{\lambda\mu}A^\nu + F^{\lambda\nu}A^\mu\,.
\end{equation}
In the same way we go on to obtain the fermionic contribution,
\begin{equation}
  J_\psi^{\lambda\mu\nu} = \frac{\partial\calL}{\partial(\partial_\lambda\psi)}
  \Delta \psi^{\mu\nu} =
  \bar{\psi}i\gamma^\lambda \bigl( x^\mu\partial^\nu - x^\nu\partial^\mu
  -i\Sigma^{\mu\nu} \bigr)\psi\,.
\end{equation}
They satisfy
$\partial_\lambda (J_A^{\lambda\mu\nu} +  J_\psi^{\lambda\mu\nu})=0$
and the conserved charge (i.e., $\lambda=0$ component) is the total angular
momentum.  From these expressions it would be a natural choice for us
to define the ``canonical'' OAM and SAM as follows;
\begin{align}
  & L_{A,\rm can}^{\mu\nu} \equiv
    -F^0_{\;\alpha}(x^\mu\partial^\nu - x^\nu\partial^\mu)A^\alpha\,,
    && S_{A,\rm can}^{\mu\nu} \equiv
    -F^{0\mu}A^\nu + F^{0\nu}A^\mu\,.\\
  & L_{\psi,\rm can}^{\mu\nu} \equiv
    i\psi^\dag (x^\mu\partial^\nu - x^\nu\partial^\mu) \psi\,,
    && S_{\psi,\rm can}^{\mu\nu} \equiv \psi^\dag \Sigma^{\mu\nu} \psi\,.
  \label{eq:canL}
\end{align}
This is simply our choice for the moment, and one may say that the
spin can be identified as the remaining operator in the homogeneous
limit where all spatial derivatives drop\footnote{The spin
  identification in such a frame to drop spatial derivatives is
  emphasized by Yoshimasa~Hidaka.  Another physical constraint is the
  commutation relation, and this prescription would always give the
  correct commutation relation of the spin.}.
These are not separately conserved quantities but only the sums, the
total angular momenta, are conserved.  We point out that the above
decomposition has been long known in the  context of the proton spin
problem (see Refs.~\cite{Leader:2013jra,Wakamatsu:2014zza} for
reviews).  In the language of quantum chromodynamics (QCD),  if the
gauge field is extended to the non-Abelian gluon field and the
temporal index is changed to $+$ in the light-cone coordinates,
$S_{\psi,\rm can}^{\mu\nu}$ and $S_{A,\rm can}^{\mu\nu}$ correspond to 
$\frac{1}{2}\Delta\Sigma$ and $\Delta G$, respectively, in what is
called the Jaffe-Manohar decomposition.

Such expressions have been known by all QCD physicists;  they look
firmly founded, but not very undoubted yet, for they are obviously
gauge dependent.  In quantum field theoreticians a common folklore is
that non-gauge-invariant objects may well be unphysical.  This story
would remind readers of a famous problem that the canonical
energy-momentum tensor is not gauge invariant, while the symmetrized
one is.  Interestingly, rotation and translational shift are coupled
together, so that the angular momenta and the energy-momentum tensor
(EMT) are linked.  The canonical EMT for the Abelian gauge theory is
derived as
\begin{equation}
  T_{A,\rm can}^{\mu\nu} = \frac{\partial\calL}{\partial(\partial_\mu A^\alpha)}
  \partial^\nu A^\alpha - g^{\mu\nu}\calL_A
  = -F^\mu_{\;\alpha} \partial^\nu A^\alpha
  + \frac{1}{4}g^{\mu\nu}F^{\alpha\beta}F_{\alpha\beta}
\end{equation}
for the gauge part, which is clearly gauge dependent, and
\begin{equation}
  T_{\psi,\rm can}^{\mu\nu} = \frac{\partial\calL}{\partial(\partial_\mu\psi)}
  \partial^\nu \psi - g^{\mu\nu}\calL_\psi
  = \bar{\psi}i\gamma^\mu\partial^\nu\psi
  -g^{\mu\nu} \bar{\psi}(i\gamma^\alpha D_\alpha - m)\psi
\end{equation}
for the fermion part.  From now on we impose onshellness and utilize
the equations of motion.  We would recall that the derivation of
N\"{o}ther's theorem already requires the equations of motion.  Then,
we can safely drop the last term in $T_{\psi,\rm can}^{\mu\nu}$ thanks
to the Dirac equation.  Then, for spatial $\mu$ and $\nu$ (denoted by
$i$ and $j$), it is straightforward to confirm the relation between
the OAM and the EMT,
\begin{equation}
  L_{A/\psi,\rm can}^{ij} = x^i T_{A/\psi,\rm can}^{0j}
  - x^j T_{A/\psi,\rm can}^{0i} \,.
  \label{eq:Lcan}
\end{equation}
So far, apart from the gauge invariance, all these relations perfectly
fit in with our intuition.

Now, let us shift gears to discussions on the symmetrized version of the
EMT{}.  To consider the physical meaning of the symmetric and the
antisymmetric parts of the EMT, the above relation~\eqref{eq:Lcan} is
quite useful.  For the gauge and the fermion parts, generally, we
immediately see that the following relation holds,
\begin{equation}
  0 = \partial_\lambda J^{\lambda\mu\nu} = \partial_\lambda
  \bigl(x^\mu T_{\rm can}^{\lambda\nu} - x^\nu T_{\rm can}^{\lambda \mu}
  + S_{\rm can}^{\lambda\mu\nu} \bigr) ~\Rightarrow~
  T_{\rm can}^{\mu\nu} - T_{\rm can}^{\nu\mu} = -\partial_\lambda
  S_{\rm can}^{\lambda\mu\nu}\,,
  \label{eq:antiT}
\end{equation}
where $T_{\rm can}^{\mu\nu}\equiv T_{A,\rm can}^{\mu\nu}+T_{\psi,\rm can}^{\mu\nu}$
and $S_{\rm can}^{\lambda\mu\nu}\equiv S_{A,\rm  can}^{\lambda\mu\nu}+S_{\psi,\rm can}^{\lambda\mu\nu}$.
Therefore, the antisymmetric part of the canonical EMT is the source
of the spin current.  The EMT as conserved currents is not unique, but
can be added by $\partial_\lambda K^{\lambda\mu\nu}$ satisfying
$K^{\lambda\mu\nu}=-K^{\mu\lambda\nu}$, which would not change the
conservation laws.  One of the most interesting and important choices
of $K^{\lambda\mu\nu}$ is,
\begin{align}
  K_{\rm Bel}^{\lambda\mu\nu} &= \frac{1}{2}\bigl( S_{\rm can}^{\lambda\mu\nu}
  -S_{\rm can}^{\mu\lambda\nu} + S_{\rm can}^{\nu\mu\lambda} \bigr) \notag\\
  &= -F^{\lambda\mu} A^\nu + \frac{i}{4} \bar{\psi}
    \bigl(-i\varepsilon^{\lambda\mu\nu\rho}\gamma_5 \gamma_\rho
    + 2g^{\mu\nu}\gamma^\lambda - 2g^{\lambda\nu} \gamma^\mu\bigr)\psi\,,
\end{align}
which gives the Belinfante-Rosenfeld form of the EMT, i.e.,
$T_{\rm Bel}^{\mu\nu} \equiv T_{\rm can}^{\mu\nu} + \partial_\lambda
K_{\rm Bel}^{\lambda\mu\nu}$.  In the above we used
$\{\gamma^\lambda,\gamma^\mu\gamma^\nu\}
=2g^{\mu\nu}\gamma^\lambda-2i\varepsilon^{\lambda\mu\nu\rho}
\gamma_5\gamma_\rho$
to reach the second line (with the conventional definition of
$\gamma_5\equiv i\gamma^0\gamma^1\gamma^2\gamma^3$).
We can show that, if $T_{\rm Bel}^{\mu\nu}$ is plugged into
Eq.~\eqref{eq:antiT}, the source is exactly canceled and
$T_{\rm Bel}^{\mu\nu}-T_{\rm Bel}^{\nu\mu} = 0$ follows, which means
that $T_{\rm Bel}^{\mu\nu}$ is symmetric.  (This is exactly the point
where many people are puzzled especially when they want to formulate
the spin hydrodynamics that seems to require antisymmetric components
of the EMT, but in this article we will not go into this issue.
Interested readers can consult a review~\cite{Florkowski:2018fap}.)

Now, we proceed to concrete expressions of the Belinfante EMT in the
Abelian gauge theory.  After several lines of calculations one can
find, for the gauge part,
\begin{equation}
  \tilde{T}_{A,\rm Bel}^{\mu\nu} = -F^\mu_{\;\alpha} F^{\nu\alpha}
  - \bar{\psi}\gamma^\mu eA^\nu \psi
  + \frac{1}{4}g^{\mu\nu} F^{\alpha\beta}F_{\alpha\beta} \,,
\end{equation}
where the second term appears from the equations of motion,
$\partial_\mu F^{\mu\nu}=\bar{\psi}i\gamma^\nu\psi$.  The fermionic
part needs a bit more labor to sort expressions out.  From the
definition it is almost instant to get,
\begin{equation}
  \tilde{T}_{\psi,\rm Bel}^{\mu\nu} = \bar{\psi} i\gamma^\mu \overleftrightarrow{\partial}^\nu
  \psi + \frac{1}{4}\varepsilon^{\mu\nu\lambda\rho}\partial_\lambda
  (\bar{\psi}\gamma_5 \gamma_\rho\psi)\,.
\end{equation}
It would be more appropriate to redefine these forms to move one term
from $\tilde{T}_{A,\rm Bel}^{\mu\nu}$ to $\tilde{T}_{\psi,\rm Bel}^{\mu\nu}$
(which unchanges the sum, i.e.,
$\tilde{T}_{A,\rm Bel}^{\mu\nu}+\tilde{T}_{\psi,\rm Bel}^{\mu\nu}
=T_{A,\rm Bel}^{\mu\nu}+T_{\psi,\rm Bel}^{\mu\nu}$), then the
gauge invariance is manifested as
\begin{align}
  T_{A,\rm Bel}^{\mu\nu} &\equiv -F^\mu_{\;\alpha} F^{\nu\alpha}
    + \frac{1}{4}g^{\mu\nu}F^{\alpha\beta}F_{\alpha\beta}\,,\\
  T_{\psi,\rm Bel}^{\mu\nu} &\equiv \bar{\psi}i\gamma^\mu
     \overleftrightarrow{D}^\nu \psi + \frac{1}{4}\varepsilon^{\mu\nu\lambda\rho}
     \partial_\lambda (\bar{\psi} \gamma_5 \gamma_\rho \psi)\,.
  \label{eq:BelTpsi}
\end{align}
These are very desirable expressions and all the terms are manifestly gauge
invariant, thus corresponding to physical observables in principle.
At this point, one might have thought that $T_{\psi,\rm Bel}^{\mu\nu}$
does not look symmetric with respect to $\mu$ and $\nu$.  In a quite
non-trivial way one can prove that the above fermionic part is
alternatively expressed as
$T_{\psi,\rm Bel}^{\mu\nu}=\bar{\psi}i\gamma^{(\mu} \overleftrightarrow{D}^{\nu)}\psi$,
which is obviously symmetric.

Coming back to the angular momentum, we can introduce the Belinfante
``improved'' form for the angular momentum, i.e.,
\begin{equation}
  J_{\rm Bel}^{\lambda\mu\nu} \equiv J^{\lambda\mu\nu}
  + \partial_\rho \bigl( x^\mu K_{\rm Bel}^{\rho\lambda\nu}
  - x^\nu K_{\rm Bel}^{\rho\lambda\mu} \bigr)\,.
\label{eq:JBeldef}
\end{equation}
Because of the antisymmetric property of $K_{\rm Bel}^{\rho\lambda\mu}$,
obviously, $\partial_\lambda J_{\rm Bel}^{\lambda\mu\nu}=0$ follows as
long as $\partial_\lambda J^{\lambda\mu\nu}=0$ holds.
Therefore, this newly defined $J_{\rm Bel}^{\lambda\mu\nu}$ may well
be qualified as a conserved physical observable.  These definitions
lead us to extremely interesting expressions, namely,
\begin{equation}
  J_{A/\psi,\rm Bel}^{\lambda\mu\nu} = x^\mu \tilde{T}_{A/\psi,\rm Bel}^{\lambda\nu}
  - x^\nu \tilde{T}_{A/\psi,\rm Bel}^{\lambda\mu} \,.
  \label{eq:JBel}
\end{equation}
Such relations imply that the total angular momentum is given by
something that looks like the OAM alone if we use the Belinfante
improved forms.  We sometimes hear people saying that the spin is
identically vanishing in the Belinfante form, but this statement
should be taken carefully.  The spin part is simply unseen and the
total angular momentum seemingly appears like the OAM even though the
spin is already included.  In the analogy to the QCD spin physics, the
angular momentum identification as in Eq.~\eqref{eq:JBel} is known as
the Ji decomposition.

\section{Dirac fermions and physical and pure gauge potentials}
\label{sec:decomposition}

Discussions on the gauge part are a little cumbersome, and in this
article we will mainly focus on the fermion part only, which, however,
does not mean we drop the gauge fields.  Let us reiterate basic
definitions from the previous overview.  In the canonical
identification, in Eq.~\eqref{eq:canL},  the OAM and the SAM are
given, respectively, by
\begin{equation}
  \bL_{\psi,\rm can} \equiv -i\psi^\dag \bx \times \bnabla \psi\,,
  \qquad
  \bS_{\psi,\rm can} \equiv -\frac{1}{2}\bar{\psi} \gamma_5 \bgamma \psi\,,
\end{equation}
where we defined $L^i\equiv\frac{1}{2}\varepsilon^{ijk}L^{jk}$ and
$S^i\equiv\frac{1}{2}\varepsilon^{ijk}S^{jk}$.  As we already discussed,
$\bL_{\psi,\rm can}$ is not gauge invariant, thus it cannot be a
physical observable supposedly.  Then, what about the Belinfante form?
We can make a decomposition using Eq.~\eqref{eq:BelTpsi}.  The latter
term may well be called the spin part, with which we can compute
$J_{\psi,\rm   Bel}^{\lambda\mu\nu}$ according to Eq.~\eqref{eq:JBel},
and subtract added terms in Eq.~\eqref{eq:JBeldef}.  Some calculations
yield,
\begin{equation}
  \tilde{\bS}_{\psi,\rm Bel} = -\frac{1}{2}\bar{\psi}\gamma_5 \bgamma\psi
  - \frac{1}{2} i \bx\times \bnabla (\psi^\dag \psi)\,.
\end{equation}
This expression is not gauge invariant, thus we shall redefine the
spin to the same form as the canonical one which is manifestly gauge
invariant and move unwanted terms to the orbital part.  Thus, in this
convention, we can reasonably adopt the following definitions,
\begin{equation}
  \bL_{\psi,\rm Bel} \equiv -i\psi^\dag \bx\times \bD \psi\,,\qquad
  \bS_{\psi,\rm Bel} \equiv \bS_{\psi,\rm can}\,.
  \label{eq:Beldef}
\end{equation}
In the high-energy physics context, the above identification is called
Ji's orbital and spin angular momentum of quarks.  Again, we make a
caution remark;  the Belinfante form has the total angular momentum
that looks like the OAM, but this does not mean that the spin
vanishes.  Some people may say that the latter in
Eq.~\eqref{eq:Beldef} cannot be true since the Belinfante EMT has no
antisymmetric part.  This kind of criticism is meaningful when we need
to construct the angular momentum in terms of the EMT, which is the
case in the spin hydrodynamics for
example~\cite{Florkowski:2018fap,Hattori:2019lfp}~\footnote{K.~F.\
  thanks Wojciech~Florkowski and Hidetoshi~Taya for simulating
  conversations on this point which seem not to be very consistent to
  each other and thus we just refer to their review and original
  literature here.}.
See also Refs.~\cite{Becattini:2011ev,Becattini:2012pp} for observable
effects of different spin tensors, which may be significant especially
in nonequilibrium~\cite{Becattini:2018duy}.  Probably one way to
define the spin part out from the Belinfante symmetrized form of the
EMT is the Gordon decomposition (as Berry defined the gauge-invariant
optical spin~\cite{Berry_2009}) which is also applicable to massless
theories.  In any case, if we do not have to refer to the EMT,
Eq.~\eqref{eq:Beldef} is just a natural way of our defining
$\bS_{\psi, \rm Bel}$, satisfying the correct commutation relation.
Now we symbolically summarize the decomposition and the corresponding
QCD terminology in Tab.~\ref{tab:summary}.

\begin{table}[t]
  \centering
  \begin{tabular}{lp{5mm}l}
    \hline
    Canonical && $\displaystyle \bJ =
                 \underbrace{-\frac{1}{2}\bar{\psi}\gamma_5 \bgamma \psi}_{\frac{1}{2}\Delta\Sigma}
                 + \underbrace{\bE \times \bA}_{\Delta G}
                 \underbrace{-i\psi^\dag (\bx\times\bnabla) \psi}_{L_{\rm can}^q}
                 + \underbrace{\bE (\bx\times\bnabla) \bA}_{L_{\rm can}^g}$ \\
    Belinfante && $\displaystyle \bJ =
                 \underbrace{-\frac{1}{2}\bar{\psi}\gamma_5 \bgamma \psi}_{\frac{1}{2}\Delta\Sigma}
                 \underbrace{-i\psi^\dag (\bx\times\bD) \psi}_{L_{\rm Ji}^q}
                  + \underbrace{\bx\times(\bE\times\bB)}_{J_{\rm Ji}^g}$
    \\ \hline
  \end{tabular}
  \caption{Breakdown of the total angular momentum $\bJ$ from
    various contributions in the canonical (Jaffe-Manohar)
    decomposition (upper) and the Belinfante (Ji) decomposition (lower).}
  \label{tab:summary}
\end{table}

Now, in this convention, the spin part has no ambiguity;  it is gauge
invariant as it should be, representing a physical observable for
sure.  The subtle (and thus interesting) point is the orbital part,
and then one may be tempted to conclude that the canonical one makes
no physical sense, and this conclusion seems to be unbreakable.  An
intriguing possibility has been suggested, however, in the high-energy
physics context~\cite{Chen:2008ag} inspired by QED studies and photon
experiments (see, for example, Ref.~\cite{Cameron_2012} for very
inspiring but a little mystical discussions including Lipkin's Zilch
which is a ``useless'' conserved charge in QED), which invoked
interesting theoretical discussions;  see Ref.~\cite{Wakamatsu:2010qj}
for example.  In fact, this canonical form can be promoted to be a
gauge-invariant canonical (gic) one (using the terminology of
Ref.~\cite{Leader:2015vwa}) as
\begin{equation}
  \bL_{\psi,\rm can} \to \bL_{\psi,\rm gic} \equiv
  -i\psi^\dag \bx\times \bD_{\rm pure} \psi\,,
\end{equation}
where $\bD_{\rm pure}\equiv \bnabla - ie\bA_{\rm pure}$.  Here, the
vector potential is decomposed into two pieces, namely,
$\bA=\bA_{\rm phys} + \bA_{\rm pure}$ with $\bA_{\rm phys}$ extracted
as a gauge invariant part and $\bA_{\rm pure}$ makes the field
strength tensor vanishing; $\bnabla\times\bA_{\rm pure}=0$.  More
specifically, under a gauge transformation, $\bA$ is changed as
$\bA\to \bA + \bnabla\alpha$, and then, by definition,
$\bA_{\rm phys}\to \bA_{\rm phys}$ and
$\bA_{\rm pure}\to \bA_{\rm pure} + \bnabla\alpha$.  One simplest
decomposition satisfying these requirements is obtained from the
Helmholtz decomposition, i.e., any vector can be represented as a sum
of divergence free (transverse) and rotation free (longitudinal)
vectors.  For a more concrete demonstration, let us write down an
explicit form as
\begin{equation}
  \bA_{\rm phys} = \bnabla \times \ba\,,\qquad
  \bA_{\rm pure} = -\bnabla \phi\,,
\end{equation}
where
\begin{align}
  & \ba(\bx) = \frac{1}{4\pi}\int_V d\bx'\, \frac{\bnabla' \times \bA(\bx')}{|\bx-\bx'|}
  - \frac{1}{4\pi}\int_S d\bS'\times \frac{\bA(\bx')}{|\bx-\bx'|} \,,\\
  & \phi(\bx) = \frac{1}{4\pi}\int_V d\bx'\, \frac{\bnabla' \cdot \bA(\bx')}{|\bx-\bx'|} 
  - \frac{1}{4\pi}\int_S d\bS'\cdot \frac{\bA(\bx')}{|\bx-\bx'|} \,.
  \label{eq:Helm}
\end{align}
In principle, now, all the terms involving $\bA$ can be made gauge
invariant.  Then, a finite difference between the canonical and the
Belinfante OAM is also a gauge invariant quantity, which is often
called the ``potential'' orbital angular momentum, i.e.,
\begin{equation}
  \bL_{\psi,\rm Bel} = \bL_{\psi,\rm gic} - e\psi^\dag \bx\times
  \bA_{\rm phys} \psi \,.
\end{equation}
Here, we make a comment which is not crucial in the present
discussions but essential for phenomenological applications and
particularly for measurability.  Even though the Helmholtz
decomposition is unique, such a gauge invariant decomposition itself
is not unique.  As discussed in Ref.~\cite{Hatta:2011zs}, for example,
a different choice could be possible and even preferable in the
high-energy processes.

We note that Eq.~\eqref{eq:Helm} is highly non-local in space, and
such ``physical'' photon should have a space-like extension.  For
static electromagnetic background fields, for example, photons are
virtual and offshell, so that space-like components are experimentally
accessible (or even the vector potentials are controlled from the
beginning).  In contrast, in the parton model at high energy, the
gauge particles are onshell and travel at the speed of light (or speed
of ``gluon'' so to speak).  Then, for such propagating modes along the
light-cone, the space-like profiles as in Eq.~\eqref{eq:Helm} are not
to be probed by scatterings.  In this case of the light-cone
propagation, as prescribed in Ref.~\cite{Hatta:2011zs}, the light-cone
decomposition would be more physical.  In the Abelian gauge theory the
alternative decomposition is as simple as
\begin{equation}
  A_{\rm phys}^i (x^-) \equiv \frac{1}{\partial^+} F^{+i}
  = \int dy^-\, \mathcal{K}(x^--y^-)\, F^{+i}(y^-)\,,
  \label{eq:Hatta}
\end{equation}
where $\mathcal{K}(x^-)$ is chosen according to the boundary
condition at $x^-=\pm\infty$ in the light-cone gauge $A^+=0$; it is
$\theta(x^-)$ for the retarded boundary condition, $-\theta(-x^-)$ for
the advanced one, and $\frac{1}{2}[\theta(x^-)-\theta(-x^-)]$ for the
mixed boundary condition.  We would point out that not only in
high-energy physics but also in the laser optics the spatially
non-local decomposition in Eq.~\eqref{eq:Helm} may not be appropriate
if the propagating lights (such as the monochromatic waves) are
concerned.  The analogy between physical contents in high-energy
physics and optics has been sometimes emphasized in the literature
(see Ref.~\cite{Leader:2015vwa} for example), but this important
question of what would be the ``natural'' choice is frequently
missing.  Along these lines of the natural choice, a mathematical
argument in connection to the geodesic in tangent space is found in
Ref.~\cite{Lorce:2012ce}.  In this article the existence of
$\bA_{\rm phys}$ suffices for our discussions at present.

\section{Potential angular momentum and physical interpretation}
\label{sec:which}

One might have a feeling that such classification of slightly
different OAMs (whilst the SAM is common in our convention) may be an
academic problem, but we recall that each term represents some
physical observable and the lack of correct understanding would cause
paradoxical confusions.  For instance, if one is interested in the
Einstein--de~Haas effect and/or the Barnett effect within a
relativistic framework, an interplay between the OAM provided by
mechanical rotation and the spin polarization measured by the
magnetization underlies observable phenomena.  We had discussed this
issue with knowledgeable researchers, some of whom told us that such a
relativistic extension of these effects may not exist after
all... such a conclusion is typically drawn based on the proper
knowledge of knowledgeable researchers that the covariant derivative
makes the theoretical formulation manifestly gauge invariant and the
derivative and the vector potential are inseparable then.  In the
previous section, however, we have already seen that we can evade this
problem by introducing $\bD_{\rm pure}$.  Now, in this section, we
would like to address a difference between $\bD$ and $\bD_{\rm pure}$.

This question would be highly reminiscent of a more familiar and
classic problem of the kinetic and the canonical momenta of a charged
particle under electromagnetic background.  That is, in our convention
of the covariant derivative, $\partial_\mu + ieA_\mu$ (i.e., $e$ is
taken to be negative), the canonical momentum should be
$\bp_{\rm can}=m\dot{\bx}+e\bA$, while the kinetic one is
$\bp_{\rm kin}=m\dot{\bx}=\bp_{\rm can}-e\bA$ in a non-relativistic
system.  Since the canonical momentum should fullfil the commutation
relation, we should identify $\bp_{\rm can}=-i\hbar\bnabla$ in the
$x$-representation and $\bp_{\rm kin}$ corresponds to the covariant
derivative.  For the gauge invariant definition of $\bp_{\rm can}$, we
can replace $\bnabla$ with $\bD_{\rm pure}$.  In other words, the
translational symmetry is generated by not the covariant derivative
but the derivative, so that $\bp_{\rm can}$ is the momentum that can
be conserved for the symmetry reason.  The difference can be easily
understood in the simplest physical example;  if a charged particle is
placed in a constant and homogeneous electric field, then the electric
field accelerates the charged particle.  Therefore, on the one hand,
$\bp_{\rm kin}$ should increase by the impulse, $e\bE t$.  On the
other hand, the vector potential $\bA=-\bE t$ gives the electric
field, and obviously, $\bp_{\rm can}=\bp_{\rm kin}+e\bA$ is time
independent and conserved.  In summary, it is important to note the
following differences:
\begin{equation}
  \begin{split}
    &\bD \quad\leftrightarrow\quad \bp_{\rm kin} \quad \text{(non-conserved)}\,,\\
    &\bD_{\rm pure}  \quad\leftrightarrow\quad \bp_{\rm can} \quad \text{(conserved)}\,.
  \end{split}
\end{equation} 
It might be little counter intuitive that $\bD$ whose definition
involves the gauge potential corresponds to the momentum carried by
the charged particle only and $\bD_{\rm pure}$ gives the total
conserved momentum.  Physically speaking, however, such a
correspondence is quite reasonable.  In most cases only particle's
$\bp_{\rm kin}$ can be directly measured and this readily measurable
quantity just corresponds to the covariant derivative.  In reality,
sometimes, $\bp_{\rm can}$ does matter as well especially when the
conservation law accounts for observable phenomena.

In exactly the same way as $\bp_{\rm kin}$ and $\bp_{\rm can}$ of the
charged particle, we can classify two orbital angular momenta as
\begin{equation}
  \begin{split}
    &\bx\times\bD \quad\leftrightarrow\quad \bL_{\rm kin}
    \;\sim\; \bL_{\psi,{\rm Bel}} \quad \text{(non-conserved)} \,,\\
    &\bx\times\bD_{\rm pure}  \quad\leftrightarrow\quad \bL_{\rm can}
    \;\sim\; \bL_{\psi,{\rm gic}} \quad \text{(conserved)} \,.
  \end{split}
\end{equation}
The difference between $\bL_{\rm kin}$ and $\bL_{\rm can}$ is often
called the ``potential'' angular momentum (see
Ref.~\cite{Wakamatsu:2017isl} for a recent analysis of this
difference).  Unlike the above trivial example of $\bp_{\rm kin}$ and
$\bp_{\rm can}$ with a constant $\bE$,
it could be often very non-trivial to imagine what physically causes
the potential angular momentum.  To see this more, armed
with these general basics, let us turn to a concrete problem now.  We
shall take a very instructive example of
Ref.~\cite{PhysRevLett.113.240404} which is entitled, ``Is the Angular
Momentum of an Electron Conserved in a Uniform Magnetic Field?'' and
this title already explains the contents by itself.  The authors of
Ref.~\cite{PhysRevLett.113.240404} considered the time evolution of
the radial width $\rho$ of an electron motion in a uniform magnetic
field $B$ using the Schr\"{o}dinger equation.  The Hamiltonian of such
a (non-relativistic) system is given by
\begin{equation}
  H = -\frac{\hbar^2 \bnabla^2}{2m} + \frac{1}{2}m\omega_L^2 \rho^2
  -i\hbar\omega_L\frac{\partial}{\partial\phi}\,,
\end{equation}
where $\omega_L=|eB|/(2m)$ (i.e., the Larmor frequency).  In classical
physics the charged particle with electric charge $e$ and mass $m$
receives the Lorentz force to make a circular rotation with the
cyclotron frequency $\omega_c=2\omega_L=|eB|/m$.  It is easy to write down the
Heisenberg equation of motion for $\langle\rho^2\rangle$ to find that
its time evolution solves as~\cite{PhysRevLett.113.240404}
\begin{equation}
  \langle\rho^2\rangle(t) = \tilde{\rho^2} + \bigl(
  \langle\rho^2\rangle(0)-\tilde{\rho^2} \bigr) \cos(\omega_c t)\,.
\end{equation}
Because the kinetic orbital angular momentum along the magnetic
direction (which is taken to be the $z$ axis, as is the convention in
the following discussions too) depends on the moment of inertia, and
the moment of inertia is a function of the radial width, they are
related to each other as
$\langle (L_{\rm kin})_z \rangle = \text{(conserved canonical OAM)}
+ m\omega_L \langle\rho^2\rangle$.  Thus, these calculations
explicitly show that $\langle \bL_{\rm kin}\rangle$ is not conserved but
has time oscillatory behavior $\propto \langle\rho^2\rangle$.  This is
an interesting observation that illustrates qualitative differences
between the classical and the quantum motions of an electron, but not
such an unexpected one;  in general case it is not $\bL_{\rm kin}$ but
$\bL_{\rm can}$ that is conserved.  The question worth thinking is
what kind of physics fills in this gap by
$m\omega_L\langle\rho^2\rangle$.

The answer is explicated in Ref.~\cite{PhysRevLett.113.240404} -- this
gap turns out to be exactly the angular momentum of the
electromagnetic field.  As we listed up in Tab.~\ref{tab:summary}, the
electromagnetic angular momentum in the Belinfante form reads,
\begin{equation}
  J_z^{\rm field} = \int d^3x\, [\bx\times(\bE\times\bB)]_z\,.
  \label{eq:Jz}
\end{equation} 
This is an integration of $\bx$ times the electromagnetic momentum
represented by the Poynting vector, which might have looked more like
the OAM, but this is the total angular momentum as we derived
in our previous discussions of this article.  As argued in
Ref.~\cite{PhysRevLett.113.240404}, if the electromagnetic fields are
static and $\bnabla\times\bE=0$ holds, this electromagnetic angular
momentum can be rewritten into a convenient form as
\begin{equation}
  J_z^{\rm field} = \int d^3x\,
  (\bnabla\cdot\bE) (\bx\times \bA_{\rm phys})_z\,.
  \label{eq:Jzsim}
\end{equation}
Here, we note that the integration by parts with
$\bB=\bnabla\times\bA=\bnabla\times\bA_{\rm phys}$ in
Eq.~\eqref{eq:Jz} would lead to an expression similar to the canonical
one in Tab.~\ref{tab:summary} but not Eq.~\eqref{eq:Jzsim}.  Only when
$\bnabla\times\bE=0$ and $\bnabla\cdot\bA_{\rm phys}=0$ (which is the
definition in the Helmholtz decomposition) both hold, we can prove the
above simplification~\eqref{eq:Jzsim}.

\begin{figure}[t]
  \centering 
  \includegraphics[width=0.3\textwidth]{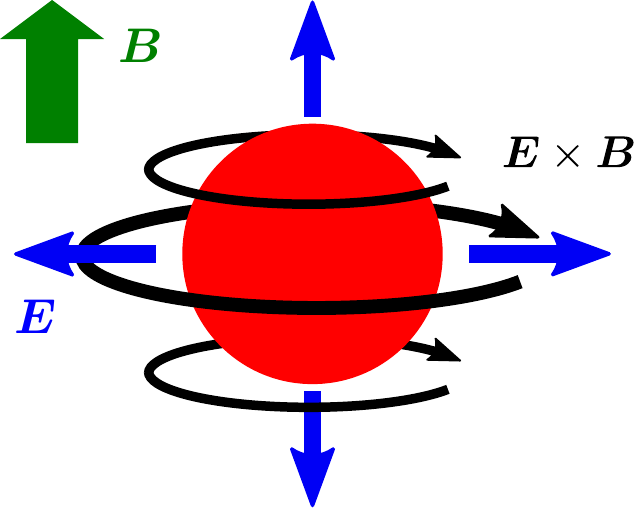}
  \caption{A charged object placed in a uniform magnetic field is 
    surrounded by the Poynting vector $\bE\times\bB$ which carries an 
    electromagnetic angular momentum contained in the conserved 
    canonical angular momentum.}
  \label{fig:charge}
\end{figure}

For a uniform magnetic field $\bA = \frac{B}{2}(-y,x,0)$ in the
symmetric gauge gives $B$ along the $z$ axis, and this already
satisfies $\nabla\cdot\bA=0$.  Then, the explicit form of
$(\bx\times \bA)_z$ is $\frac{B}{2}\rho^2$ with $\rho^2=x^2+y^2$.
Since $\bnabla\cdot\bE$ is nothing but the electric charge density,
Eq.~\eqref{eq:Jzsim} under a uniform magnetic field eventually
becomes,
\begin{equation}
  J_z^{\rm field} = \frac{eB}{2}\langle\rho^2\rangle
  = m\omega_L \langle\rho^2\rangle\,.
  \label{eq:JB}
\end{equation}
This is precisely the potential angular momentum!  There is a plain
explanation of why $J_z^{\rm field}$ should appear to make the
conserved angular momentum.  Figure~\ref{fig:charge} is a
corresponding illustration of a charged object placed in a uniform
magnetic field.  The red blob represents a charged particle
distribution (i.e., charge density in classical physics and
probability distribution in quantum mechanics).  Such a charged object
is a source resulting in Coulomb electric fields $\bE$, and
$\bE\times\bB$ goes around the charged object.  In this illustration
the charge is taken to be positive, but for an electron as we assumed
in this section, the electric field should be directed oppositely and
the Poynting vector goes in the other way around.  Because of this
circular structure of the Poynting vector, the electromagnetic fields
have a nonzero angular momentum, which was found to be
Eq.~\eqref{eq:JB}.

Still, the physical interpretation is quite non-trivial, we must say.
Literally speaking, $J_z^{\rm field}$ is a purely electromagnetic
contribution, and nevertheless, $\bE$ extends from the charge source
and in this sense we may well say that $\bE$ is rather attributed to
the matter property.  If we are interested in the mechanical rotation
as is the case in the Barnett and the Einstein--de~Haas effects,
however, we should count the kinetic angular momentum.  Even in that
case, this extra electromagnetic contribution could affect the kinetic
angular momentum through the angular momentum conservation law.

\section{Feynman's angular momentum paradox and possible relevance to the relativistic nucleus-nucleus collision}
\label{sec:feynman}

Careful readers might have realized that the argument about
$J_z^{\rm field}$ is essentially rooted in Feynman's angular momentum
paradox in classical physics.  The paradox is articulated in
\textit{The Feynman Lectures} and the original setup is composed from
a conductor disk with a solenoid that controls the magnetic strength.
For detailed analysis of the original version of Feynman's angular
momentum paradox, see Ref.~\cite{Fparadox} for example.  Here, let us
discuss a simplified version of Feynman's angular momentum paradox.

\begin{figure}[t]
  \centering 
  \includegraphics[width=0.3\textwidth]{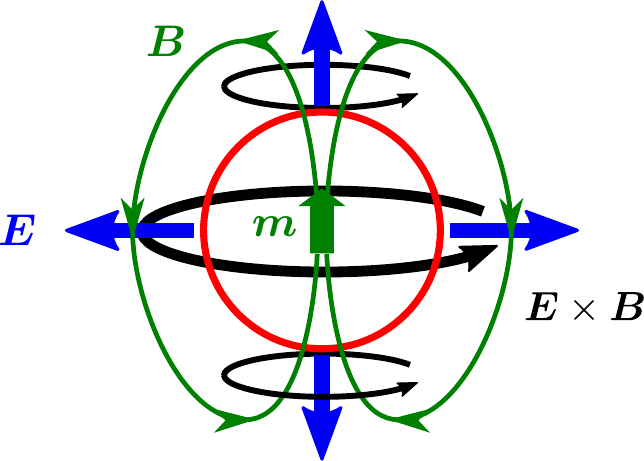}
  \caption{A charged thin sphere (red circle) and a magnetic moment at the center of the 
    sphere.  The dipolar magnetic fields and the Coulomb electric
    fields make circulating Poynting vectors.}
  \label{fig:sphere}
\end{figure}

We suppose that a thin sphere is uniformly charged (whose total amount
is denoted by $Q$) and a finite magnetic moment $\bbm$ is fixed at the
center of the sphere (see Fig.~\ref{fig:sphere}).  The electric
(outside of the sphere) and the magnetic profiles are, respectively,
\begin{equation}
  \bE = \frac{Q}{4\pi} \frac{\bx}{r^3}\,,\qquad
  \bB = \frac{1}{4\pi r^3} \biggl( \frac{3\bbm\cdot\bx \bx}{r^2}
  -\bbm \biggr)\,.
  \label{eq:EBsphere}
\end{equation}
If $\bbm$ changes as a function of time, the magnetic field changes as 
well, which also results in an induction electric field due to
Amp\`{e}re's law.  Then, the charged sphere feels a moment of force
under this induced electric field, $\bE_{\rm ind}$, and the sphere is
accelerated for rotation.  The space integrated moment of force is,
after some patient calculations, found to take a form of
\begin{equation}
  \boldsymbol{N} = \int d\boldsymbol{S}\cdot 
  \frac{\bx\times Q\bE_{\rm ind}}{4\pi R^2} = -\frac{Q \dot{\bbm}}{6\pi R}\,,
\end{equation}
where $R$ denotes the radius of the sphere.  Therefore, if $\bbm$
decreases, the sphere takes a positive moment of force to acquire a
mechanical angular momentum.  The question is; how can the angular
momentum conservation law be satisfied?  This phenomenon may sound
similar to the Einstein--de~Haas effect, but one should recall two
important differences.  One is that the object should be charge
neutral in the Einstein--de~Haas effect, and another is that in this
classical example there is no magnetization at all.  There are many
variants of Feynman's paradox, and they usually belong to classical
physics (no spin effects).

Readers should be already aware of the resolution.  As indicated in
Fig.~\ref{fig:sphere}, the electromagnetic field generates circulating
Poynting vectors.  Actually, from explicit expressions of
Eq.~\eqref{eq:EBsphere}, we can obtain the angular momentum
distribution as
\begin{equation}
  \bx\times (\bE\times\bB) = \frac{Q}{(4\pi)^2 r^6}(r^2 \bbm
  -\bx \bx\cdot\bbm)\,.
\end{equation}
Therefore, the total angular momentum integrated in space outside of
the sphere turns out to be,
\begin{equation}
  \bJ^{\rm field} = \frac{Q \bbm}{6\pi R}\,.
\label{eq:dipoleJ}
\end{equation}
It is obvious that the angular momentum in mechanical rotation
originates from the loss in $\bJ^{\rm field}$, so that the total
angular momentum is surely conserved.  See
Ref.~\cite{doi:10.1119/1.15597} for related discussions on the
Poynting vector contributions in classical electromagnetism.
Interestingly this result of Eq.~\eqref{eq:dipoleJ} was extended to
the one-loop QED level which turned out to be free from a
short-distance cutoff~\cite{DAMSKI2019114828}.

In this classical example of Feynman's paradox the essential point is
that either $\bE$ or $\bB$ changes to make a finite difference in
$\bx\times(\bE\times\bB)$ from which the mechanical rotation is
induced.  The novelty in the quantum mechanical example seen in the
previous section is that quantum oscillations exhibit time dependence
even for constant $\bE$ and $\bB$.  In both cases the important lesson
is that, as long as we prefer to use the Belinfante improved form for
the EMT and the angular momenta, the covariant derivative in the
matter sector makes all the expressions manifestly gauge invariant,
and then we can access the kinetic angular momentum of the matter
which is not necessarily conserved.

\begin{figure}[t]
  \centering 
  \includegraphics[width=0.3\textwidth]{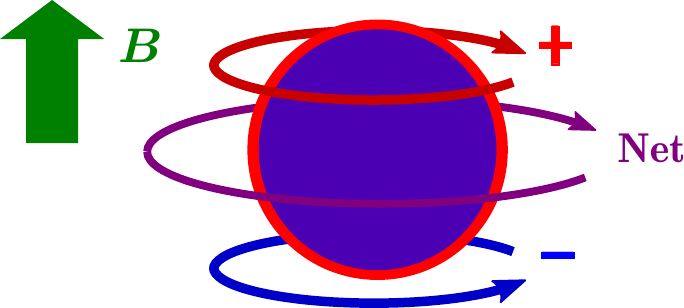}
  \caption{A net induced angular momentum with faster rotating
    positively and negatively charged particles.  There are more
    positively charged particles in a plasma because of protons in the
    participant particles.}
  \label{fig:net}
\end{figure}

So far, we have been having general discussions not specifying any
experimental realizations at all.  Let us now consider some possible
applications to the high-energy nucleus-nucleus collisions.  It is
known that the OAM in the non-central nucleus-nucleus collision can
reach a gigantic value as large as $\sim 10^5\hbar$ as evaluated in
the AMPT model~\cite{Jiang:2016woz}, supported by experimental
data~\cite{STAR:2017ckg}.  Here, we can make an order of
magnitude estimate of extra angular momentum from the decay of the
magnetic field using Eq.~\eqref{eq:JB}.  Our following discussions may
look different from Ref.~\cite{Guo:2019mgh} which addresses a
possibility of the spin polarization by the induced electric fields.
There are some discrepancies from spatial inhomogeneity as well as
temporally decaying magnetic properties and also from hydrodynamic
treatments, but we note that microscopically underlying physics is
common.

The magnetic field created right after the collision is of order
$eB\sim \text{GeV}^2$ at largest, and $\langle\rho^2\rangle$ in the
collision geometry is around $\sim 10\,\text{fm}^2$.  Therefore, if
the magnetic field quickly decays whose time scale is
$\sim 0.1\,\text{fm/c}$, this field angular momentum,
$J_z^{\rm field}\sim 10\,\text{GeV}^2\!\cdot\!\text{fm}^2 \sim 100\hbar$,
is transferred to the angular momentum of single particle.  The net
charge is $0.1Z\sim Z$ depending on the impact parameter and the
baryon stopping, where $Z\sim 100$ is the atomic number of the heavy
nucleus, and so the net angular momentum is of order
$10^3\sim 10^4\hbar$.  Here, we would emphasize that the time scale is
irrelevant.  This angular momentum arises as a consequence of the
conservation law and it is just there for any fast decaying $B$
(except loss by polarized photon emissions).  From this simple
estimate we can conclude that the net induced angular momentum is
significantly smaller than the primarily produced angular momentum
$\sim 10^5\hbar$.  This is, however, not yet the end of the story.  In
reality of the nucleus-nucleus collision a plasma state consists of
positively and negatively charged particles and the net charge is only
its small fraction.  Then, we can anticipate at least an order of
magnitude larger angular momenta for positively and negatively charged
components in the opposite directions which mostly cancel to lead to
the net angular momentum (see Fig.~\ref{fig:net}).  If this
two-component model is a good approximation (which is dictated by the
interaction strength between two components), each charged sector
could carry the induced angular momentum $\sim 10^4\sim 10^5\hbar$,
comparable to the primarily produced angular momentum.  Interestingly,
such a two-component picture with opposite rotation has been confirmed
in the numerical simulation for the Einstein--de~Haas effect in cold
atomic systems~\cite{PhysRevLett.96.080405,ebling2017einsteinde}.

We have some more ideas [say, the global polarization should be also
associated with the field angular momentum by Eq.~\eqref{eq:dipoleJ}
whose effect has never been studied] and have in mind applications to
the local polarization measurements, but we shall stop our stories
here.  Such ideas as well as more detailed and quantitative
calculations will be reported in a separate publication.

\section{Epilogue}
\label{sec:epilogue}

The interplay between the OAM and SAM is an old subject, but its
entanglement with chirality in a relativistic framework is a quite new
research field.  The ultra-relativistic nucleus-nucleus collision
experiments have been offering inspiring data and high-energy nuclear
physicists have become wiser and wiser over decades.  Some people,
especially researchers close to but not directly in our field, might
have assumed that physics of the relativistic nucleus-nucleus
collision passed a peak.  We must say, such an assumption is nothing
but a hasty conclusion.  The nucleus-nucleus collision still continues
to provide us with surprises one after another.

Recent investigations on the OAM and SAM decomposition and their
interactions are motivated by the $\Lambda$ and $\bar{\Lambda}$
polarization measurements, but we should emphasize that this is not a
hip excitement.  Theoretically speaking, this is an extremely profound
subject, and there are still many things that nobody has understood.
One common criticism against such kind of theory problem would be;
what you call ``profound'' is just what I would call ``academic'', or
give me any measurable observable?  Indeed it is not easy to make a
new proposal for the nucleus-nucleus collision.  Nevertheless, we can
export our ideas inspired by the nucleus-nucleus collision to other
physics fields such as cold atomic systems and laser optics.  Still,
even if exported ideas are adapted in a different shape, we can
proudly say that this is a tremendous achievement from the high-energy
nuclear physics!

We also emphasize that the OAM/SAM decomposition and also the EMT
measurements are of central interest of the future coming electron-ion
collider (EIC) physics.  At least three pretty independent
communities, the heavy-ion collision, the proton spin, and the laser
optics have worked on the very similar physics, and now is the time to
put all our wisdoms together toward the next generation breakthrough.
\vspace{0.3em}

We would like to make acknowledgments.  We thank Zebin~Qiu for
successful collaborations.  K.F.\ is grateful to Kazuya~Mameda for
extremely useful discussions about ongoing projects on the
Einstein--de~Haas effect.  K.F.\ also thanks Yoshi~Hatta for
interesting and critical (as always) conversations.  K.F.\ also would
like to acknowledge very useful and sometimes confusing (and so
interesting) conversations with Francesco~Becattini,
Wojciech~Florkowski, and Xu-Guang~Huang.

\bibliographystyle{utphys}
\bibliography{chiral}

\end{document}